\documentstyle[prb,aps,floats,graphicx]{revtex}
\draft

\newcommand{{ \vn }}{\vec{n}}

\newcommand{\beq}{\begin{equation}}
\newcommand{\eneq}{\end{equation}}

\newcommand{\met }{\frac{1}{2}}
\newcommand{\meV}{\mathrm{meV}}
\newcommand{\mK}{\mathrm{mK}}

\begin{document}
\draft
\preprint{jouault et al}
\twocolumn[\hsize\textwidth\columnwidth\hsize\csname@twocolumnfalse\endcsname 
\title{Sequential magnetotunneling  in  a vertical  Quantum Dot tuned at 
the  crossing to  higher 
spin states }

\author{B. Jouault}
\address{
INFM Unit\'a di Napoli, Mostra d'Oltremare, Pad 20, 80125 Napoli (Italy) }
\address{L2M, CNRS,Bagneux,(France)}
\author{G. Santoro}
\address{INFM and Scuola Internazionale Superiore di Studi Avanzati (SISSA),  Via Beirut 4, 34013 Trieste (Italy)} 
\author{A. Tagliacozzo}
\address{
INFM Unit\'a di Napoli, Mostra d'Oltremare, Pad 20, 80125 Napoli (Italy) }
\address{Institut f\"ur Nanotechnologie (INT), Forschungszentrum, Postfach 3640, D-76021 Karlsruhe (Germany)} 

\maketitle
\begin{abstract}
We have calculated the 
linear magnetoconductance  across a vertical parabolic Quantum Dot
with a magnetic field in the direction of the current. Gate voltage 
and magnetic field are tuned 
at the degeneracy point between the occupancies $N=2$ and $N=3$,
 close to 
the Singlet-Triplet transition  for  $N=2$.
We find that the conductance is 
enhanced prior to the transition 
by nearby crossings of the levels of the $3$ particle dot. Immediately 
after it is depressed   by roughly $1/3$,  
as long as the total spin $S$ of the 3 electron ground state doesn't 
change  from $S=1/2$ to $S=3/2$,  due to spin selection rule. 
 At low temperature this dip is very sharp, but the peak is recovered by
increasing the temperature. 
\end{abstract}


]

\section{Introduction}
\begin{figure}
\centering
\includegraphics[width=\columnwidth]{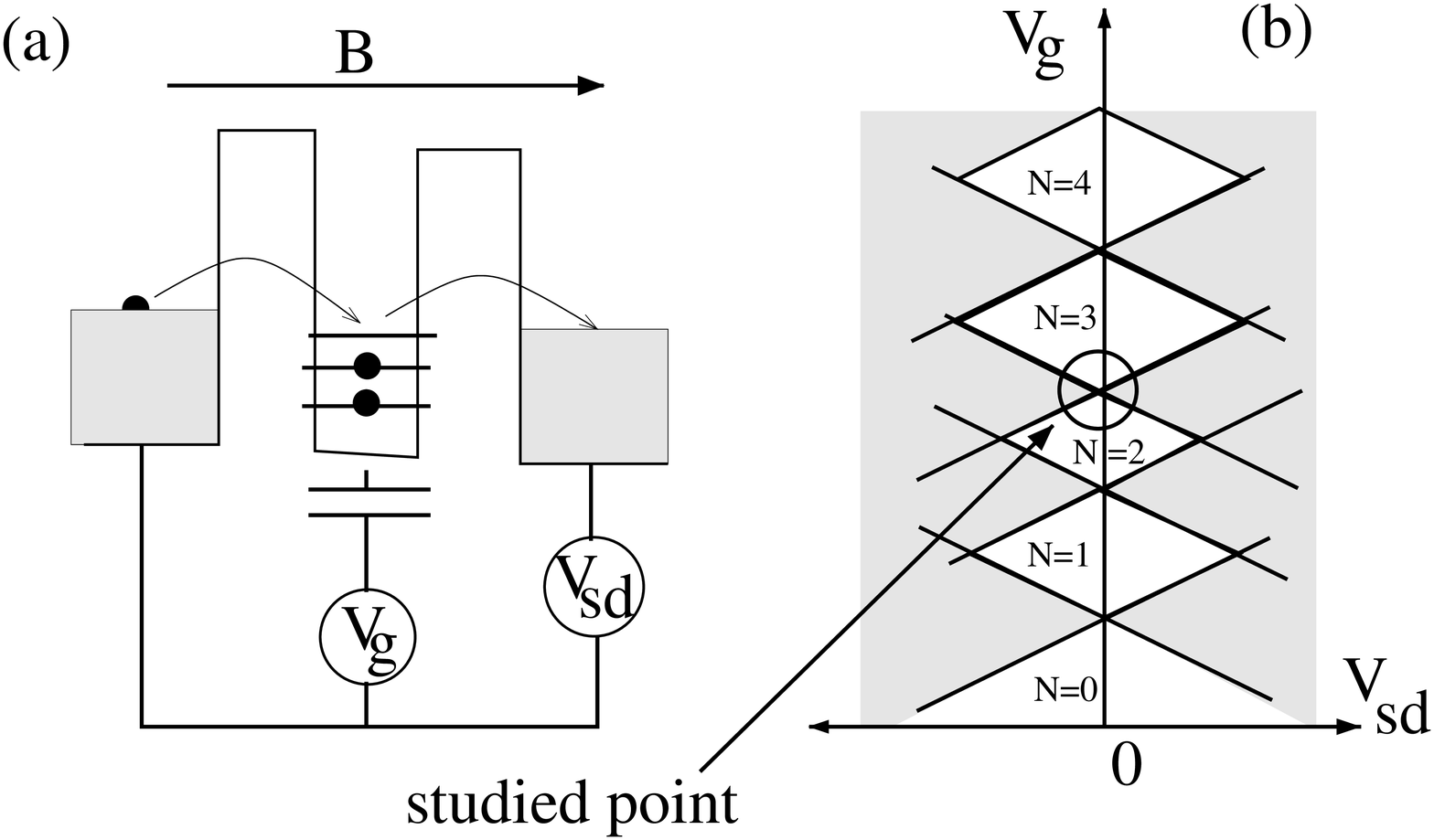}
\caption{({\bf a})Sketch of the conduction band of the device in a 
linear transport experiment with a magnetic field B along the 
direction of the current. ({\bf b}) Grey-scale conductance plot in the plane $(V_g,V_{sd})$. The circle  encloses
the conductance peak at the degeneracy point $E^{(N=3)}_0 (V_g,B)=E^{(N=2)}_0 
(V_g,B) $ studied in this work.}
\label{scheme}
\end{figure}

The conductance of a clean vertical Quantum Dot  (QD) versus a gate voltage 
$V_g $ and  a source-drain voltage $V_{sd}$ allows the detailed study of the 
ground state (GS) and first excited states of few electrons confined in the 
dot~\cite{tarucha,kouwen,kouwen2,weis} (see fig.~\ref{scheme}).
 The discrete levels due to the parabolic 
confining potential introduce a shell structure in the electron filling.
In the case of a circular disk,
application of a magnetic field $B$ in the direction of the current ($z$-direction), 
orthogonal to the dot, favours states with increasing angular momenta, 
as well as higher total spin, also 
in the absence of  Zeeman spin splitting which is believed to be small 
in these systems~\cite{maksym-yang}. The location  of the crossings 
depends on the ratio between the Coulomb energy scale $U$ 
and the level spacing of the 
confinement potential $\hbar \omega _d $. We consider here the case 
$U \sim \hbar \omega_d$. For larger ratios,
transitions with $B$ to higher spin states in dots 
with few electrons have  been  related to an ``electron molecule''
picture for the particle distribution within the dot~\cite{aoki-grabert}.

These features are strongly dependent on 
the long ranged interaction. 
Therefore,  exactly soluble models like the Constant 
Interaction (CI) model  or even the one with harmonic e-e interaction 
(HI)
are unable to reproduce them~\cite{payne}.
In fact, the  HI model  allows only for ground 
states which are  singlet (S) for $N$ even ($N$ is the number of 
electrons  in the isolated dot) or doublet (D) for $N$ odd.
 On the contrary, it was long ago recognized that a two-electron 
dot undergoes a Singlet-Triplet (S-T) transition when the magnetic field is 
large enough~\cite{wagner}.  This was indeed seen~\cite{kouwen} at 
$B_{ST}\sim 4$~Tesla in dots whose 
confining energy $\hbar \omega_d$ is about 5~meV.
In fact, at difference with  what happens  in the $^4He$ atom
 ($B_{ST}\sim 10^5$Tesla), here the S-T energy separation 
at $B=0$ is only $\sim$ 5~meV. Doublet-Quadruplet (D-Q) transitions 
for an $N=3$ particle dot  
are correspondingly found~\cite{pfann} (see below). 
Effects on the conductance are expected which are usually named  
``spin blockade'' phenomena\cite{weinmann}:

$a)$  if the total spin 
of the ground state of the $N+1$ and $N$ particle dot
($^{N+1}\!GS, ^{N}\!GS$ in the following)  differ by more than $1/2$, the
 dot is blocked, with the corresponding peak in the linear conductance missing 
at zero temperature.

 $b)$ The reduction of the total spin in adding an extra electron to
the dot has a lower transition rate as compared to spin increase.
This can lead to negative differential conductance.

In this work we analyze in detail the behavior of the linear conductance
of the dot when the gate voltage is such that the GS energies
for $N=2$ and $N=3$  are degenerate 
(see fig.~\ref{scheme}{\bf b}) and 
we show that, by increasing  $B$ beyond 
the S-T transition of the $N=2 $ particle dot, the conductance displays 
a sudden dip  due to the spin selection rule $b)$.
 Because  the corresponding 
$^3\!GS$  has total spin $1/2$, 
the conductance is  depressed  by roughly $1/3$.
 In fact, due to conservation of $S_z$,
it is impossible to add  an electron with spin up (down)
to the $N=2$ particle dot in the triplet state with $S_z =1 (-1)$, to give
total spin $S=1/2$. These two (over six) possibilities that are missing
are restored when
 the  $^3\!GS$  changes  to $ S=3/2$.    
 At low temperatures this dip is very sharp, but the peak is recovered in
increasing the temperature. 

 Three main ingredients are required to calculate the linear conductance, 
according to second order perturbation theory in the tunneling:

$i)$  the energies $^N\!E_{\alpha}  $ of the isolated dot
with e-e interactions (where N=2,3 and $\alpha$ labels the states of 
the dot).

$ii)$  the spectral amplitude $Z^{N+1,\alpha}$ for electron addition to 
the dot.

$iii)$ the one-particle tunneling matrix $\Gamma$  from the 
contacts to the dot.

The first two quantities are provided by exact diagonalization of the 
isolated dot, which uses a Lanczos code (up to 7 electrons), already employed 
previously in the analysis of the magnetoconductance data in pillar 
structures~\cite{faini}. The third quantity was  calculated 
analytically~\cite{benoit}.
In Section II we study  the transition rates in a dot from $N=2$ to 
$N=3$ by addition of an extra electron in an orbital of given 
quantum numbers.
 We analyze the various contributions to the tunneling,
and the  details  of the spectral amplitude around $B_{ST}$.
We find that the  spin  selection rule mentioned above is responsible for 
a marked difference  in the peak heights for transitions from 
${^2T^1} \to {^3D^{1,2}} $ and ${^2T^1} \to {^3Q^3}$ (see table I for 
the notation). 
The linear magnetoconductance
 displays  a sharp dip at low temperature  
at the S-T transition. 
In Section III we comment the results: 
similar peculiarities of the energy spectrum around $B_{ST}$ can be found 
for larger $N$ ($N > 3$). They 
 are remnant of the S-T transition in the $N=2$ particle dot. 
However, in increasing $N$, the number of crossings is  expected to increase 
 and they  occur in a smaller energy range, so that their effect on the 
transition rate could be averaged out by temperature. 

\section{Linear magnetoconductance at the S-T transition}

Transitions of an isolated dot to higher spin states with increasing 
magnetic field, follow from an interplay between  the 
e-e interaction and the confining potential. They are easily monitored 
with exact diagonalization of the electron system. In fig.~2  the position 
in energy of the first levels of a two-dimensional(2D) QD for  $N=2$ and $N=3$
is plotted versus magnetic field, expressed in $\hbar\omega_c$.
Here $\omega_c = eB/m^*c $ is the cyclotron frequency and
$m^* =0.067 m_e $ is the  effective electron mass in $Ga As$.
States are labeled by the total spin $S$ and by the components  
(along the direction of the field $B$) of the total spin
$S_z$ and of the total angular momentum $M$ (see Table I). 

The dot is laterally 
confined by a parabolic 
potential with level spacing  $\hbar\omega_d=4$~meV. The e-e interaction is 
$U \times l_d/|\mathbf{r}-\mathbf{r}'|$, where 
$l_d = \sqrt {\hbar/m^*\omega_d}$ is the  length  scale
 and the vectors are in
the dot plane.
 The energy scale of the 
 Coulomb interaction  $U$ is taken equal to 3~meV.
Spin splitting is not included in the picture for clarity. 
It is $ g^* \mu_B B \sim$ 0.1~meV ($\mu_B$ is the Bohr magneton)
at these fields  if the bulk gyromagnetic 
ratio for $GaAs$ ($g^*=-0.44$)  is assumed, but it could be 
lower\cite{dobers}.
A Lanczos  algorithm is used on a basis which includes Slater 
determinants  constructed with up to $28$ single particle orbitals
 that are solutions of the 2D  harmonic confining potential.
\begin{figure}[!b]
\centering
\includegraphics*[width=\columnwidth]{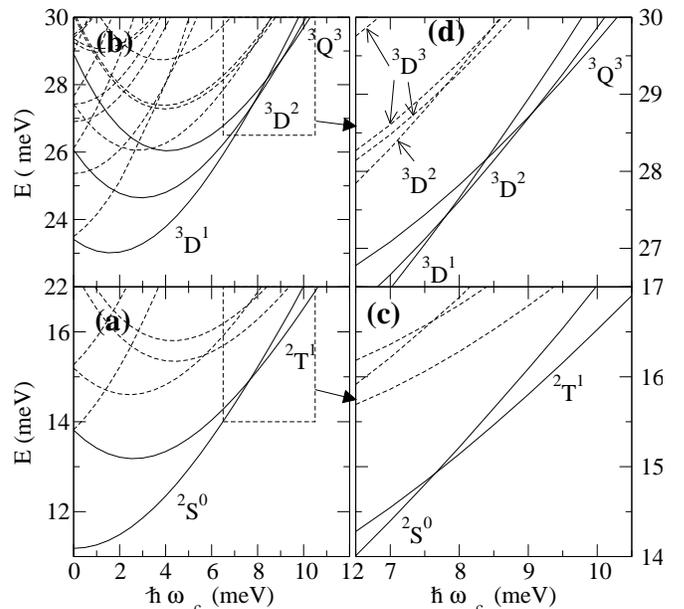}
\caption{Energy levels of an isolated  dot vs $\hbar\omega_c$. 
Zeeman spin splitting 
is not included. ({\bf a}) and 
({\bf c}): 
dot with  2 electrons. ({\bf b}) and ({\bf d}): dot with  3 electrons.
Fig.s ({\bf c}) and ({\bf d}) magnify  the rectangle areas of ({\bf a}) 
and ({\bf b})
respectively. The level spacing of the confining 
potential is $\hbar\omega_d = $ 4~meV, the strength of the Coulomb interaction 
is  $U =$ 3~meV.  The GS 
levels $^2S$,$^2T$,$^3D^1$,$^3D^2$,$^3Q^3$ are represented by solid lines
(notation in Table I).
 Higher excited levels are represented by dashed lines. }
\label{crossing}
\end{figure}

The  S-T crossing in the $ ^2GS$  (fig.~2{\bf a,c}) 
occurs at $\hbar\omega_c
\sim $ 7.5~meV.
The next available levels are also shown and are located rather 
higher in energy.
Correspondingly, the $^3GS$ (fig.~2{\bf b,d})
 is the  doublet $^3D^1$ at lower $\hbar\omega_c$ and becomes the 
 doublet $^3D^2$ at  $\hbar \omega _c =$  7.8~meV, 
according to the general rule that higher magnetic fields favor larger
angular momenta. Next, there is  a doublet-quadruplet (D-Q)
transition at the magnetic field $B_{DQ}$ 
($\hbar\omega_c \sim 9$~meV). Therefore, the sequence of 
transitions with increasing  $B$, for addition of one electron at the 
lowest energy cost, is the following: 
$^2S \rightarrow  {^3D^1}$,
$^2T \rightarrow {^3D^1}$,
$^2T \rightarrow {^3D^2}$,
$^2T \rightarrow {^3Q ^3}$.
The ${^3D ^1}\to{^3D^2}$ 
transition  always occurs at a higher magnetic field than the S-T transition
in the  $^2GS$: this has been numerically  verified by varying 
the confinement of the dot.

We calculate the transition rates for addition of one extra electron ($nm\sigma$) from the contacts.
$n$ and $m$ are the principal quantum number and the angular momentum
which label the one-particle energies $ \epsilon_{ n m}= 
\hbar \omega_0 (n  +1 )+ m \hbar \omega_c/2$  ($\omega _0
= \sqrt { \omega_d^2 +  \omega_c^2 /4}$). Here  $m$ has the same parity 
as $n$,  taking the values $ -n , -n+2, ... , n-2 ,n $.

\begin{figure}[!t]
\centering
\includegraphics*[angle=-90,width=\columnwidth]{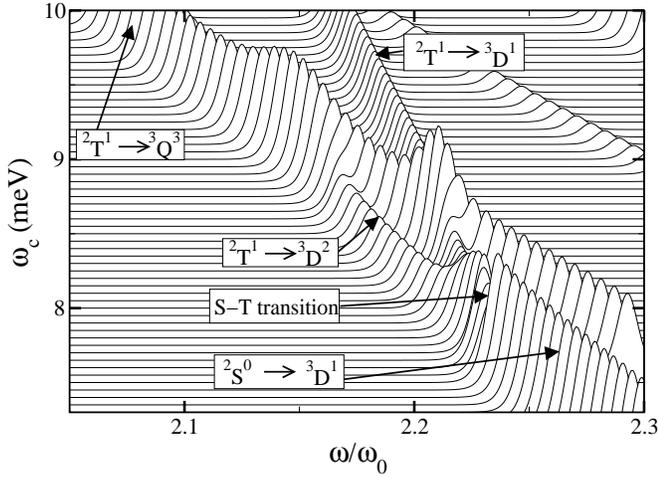}
\caption{Transition rates  as a function of $\omega/\omega_0$ where
 $\hbar\omega= E-{^2E_0}$ is the energy difference with respect to the
 $^2GS$ energy  ($\omega _0
= \sqrt { \omega_d^2 +  \omega_c^2 /4}$). Zeeman spin splitting is not 
included. The step 
in $\hbar \omega_c$ is equal to 0.05~meV.
 Curves are shifted for clarity. A thermal broadening 
 of $ T \sim 200\mK$ has been added.
  The dot parameters are  $\hbar \omega_d =4~\meV, U=3~\meV$.}
\label{3dfig}
\end{figure} 

The transition rate reads:
\begin{eqnarray}
t_{N,0}^{N+1, \alpha }(\epsilon )= \sum _{nm,n'm'} \Gamma _{nm}^{n'm'}
 (\epsilon) 
Z_{n m \sigma}^{N+1,\alpha} \left [ Z_{n' m' \sigma}^{N+1,\alpha} 
\right ]^* \nonumber \\
 \delta \left ( \epsilon  - (E_{\alpha}^{N+1}- E_0^N)\right ). 
\label{trans} 
\end{eqnarray}
Here $ Z_{n m \sigma}^{N+1,\alpha} 
= \left < N+1, \alpha \right .\left |
c^{\dagger}_{nm\sigma}\right | N,0\left .\right >$
  is the spectral weight amplitude.
The dot is assumed to be in the $^NGS$, $|N 0 \rangle $,
 prior to electron addition and 
  $\alpha \equiv (S,M)$ labels the final state with $N+1$ 
electrons. 

The matrix $ \Gamma _{nm}^{n'm'}$ appearing in eq.~\ref{trans}
 contains  the single particle 
orbital overlaps. It describes the coupling of the dot to the contacts
which have highly degenerate Landau subbands and is 
factorized in the $z$ and orthogonal directions. It is also diagonal in the 
$m$ indices, as
 the component of the 
angular momentum in the $z-$ direction is conserved in the tunneling due to 
the assumed cylindrical symmetry.
 Peaks in the transition rate of eq.~\ref{trans} 
are present each time  the energy $\epsilon$ matches  
$\mu^{N+1,\alpha}_{N,0} \equiv \:{^{N+1}E_\alpha} - {^NE_0}$, 
provided conservation of
 $S,S_z$ and $M$ are assured when adding the extra electron.
We neglect Zeeman spin 
splitting: its effect  will be discussed shortly 
in the conclusion.
The transition rates are 
plotted in fig.~\ref{3dfig}
 {\it vs} the dimensionless ratio 
$\omega/\omega_0$  for increasing magnetic field. Here 
 $\hbar \omega= E-{^2E _0}$ is the excitation energy with respect to the 
$^2GS$.
Curves (at intervals of 
$\Delta \hbar \omega_c = 0.05~\meV$) are shifted upwards for clarity 
with increasing $ \omega_c$. An extra  width
has been given to  the peaks (which are $\delta -$like in eq.~\ref{trans}),
what is consistent with a thermal broadening of $200mK$
(see below). Each series of peaks 
corresponds  to the addition of one particle of definite angular momentum
to the $^2GS$ . 
As an example, the peak from the singlet GS  at 
$\omega/\omega_0 \approx 2.3 $, for the lowest magnetic field,
corresponds to the addition
of one electron with $m=1$, what 
changes the state of the dot from $^2S^0$ to $^3D^1$.
The initial and final states of the 
transition are labeled within the figure. 
The landscape changes drastically when the $^2GS$ becomes a triplet. 

In the plot of fig.~\ref{3dfig} only the diagonal term $ n=n'$ has been 
included 
in the sums of  eq.~\ref{trans}. 
This is expected to give the largest contribution 
within  weak coupling~\cite{sun}. 
The spectral weight  $\left |Z_{n m \sigma}^{N+1,\alpha}\right |^2$ is largely 
responsible for the heights of the peaks in fig.~\ref{3dfig}. 


\begin{figure}[!b]
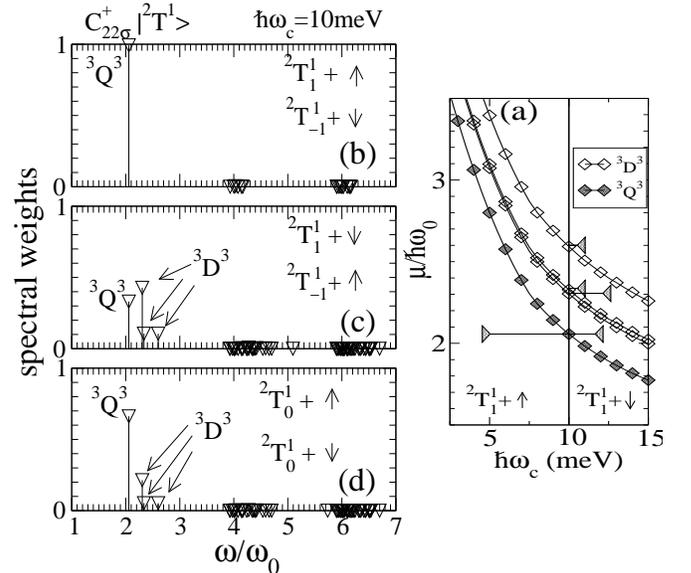

\begin{minipage}{0.6 \columnwidth}
\includegraphics*[width=\textwidth]{./green3.eps}
\end{minipage}%
\begin{minipage}{0.4 \columnwidth}
\includegraphics*[width=1\textwidth,height=2in]{./gl3.eps}
\end{minipage}%
\caption{ ({\bf a}): The energy levels with $M=3$ vs
 $\hbar \omega_c$ are plotted on the right side. The spectral weights 
are depicted as horizontal lines to the right (+$\downarrow$) 
and to the left(+$\uparrow$) of the vertical line.
({\bf b}),({\bf c}),({\bf d}): Spectral weights for the ${^2T^1} 
\rightarrow {^3Q^3}$ transition at  $\hbar \omega_c$=10.0~meV.
 $\hbar \omega = (E- {^2E_0} )$ 
are the excitation energies.
Dot  parameters as in fig.~3. }
\label{green3}
\end{figure}

This is seen in  fig.~\ref{green3}{\bf a,b,c} for 
 the case when an electron $n=2,m=2$ is added to the $^2T^1_{1}$
 state (denoted symbolically as 
$ C^{\dagger}_{22\sigma}|{^2T^1_1}>$),
 producing the final state with $M=3$. 
There are four low lying states  in the N=3 energy spectrum, labeled 
$^3Q^3$ and $^3D^3$  (fig.~\ref{green3}{\bf a}).
 The lowest is the quadruplet state $^3Q^3$.  The 
spectral weight  for addition
of an electron with spin down  is distributed among them 
($r.h.s.$ of fig.~\ref{green3}{\bf a} and fig.~\ref{green3}{\bf c}),
while in the case of  a spin up  particle 
the spectral weight  is practically  exhausted  by the  transition 
to the quadruplet state
with $S_z =3/2$ ($l.h.s.$ of fig.~\ref{green3}{\bf a} and 
 fig.~\ref{green3}{\bf b}). Up and down arrows are 
to be interchanged in the case of $^2T^1_{-1}$~\cite{nota}.         
Transition rates are proportional to the square  of the 
Clebsch-Gordan coefficients
$\left| \langle S' S_z' | S S_z \met \pm \met \rangle_{CG} \right|^2$ 
which accounts for the
spin selection rules~\cite{weinmann}.

These results are fully obvious in a picture in which Coulomb interaction
is treated as a perturbation. 

Starting by filling single particle levels  of the parabolic confinement
potential, the $N=3$ state  (lowest in energy) with  $S_z=1/2$ and $M=3$
is   obtained from the linear 
combination of three degenerate states  with two electrons up  and one 
down, distributed  between $n=0,m=0; n=1,m=1;n=2,m=2 $. Coulomb 
interaction breaks the degeneracy, giving rise  to the quadruplet state 
 $|{^3Q_{1/2}^3}\rangle$,
 which is the properly antisymmetrized combination of the 
previous three (with normalizing factor $1/\sqrt{3}$) and two doublet 
states ${^3D_{1/2}^3}$, 
which are almost unsplitted (see fig.~\ref{green3}{\bf a}).
The third doublet state at higher energy in 
 fig.~\ref{green3}{\bf a} is obtained by placing one electron
in $n=3,m=3$ and the two others in $n=0,m=0$.
The transition ${^2T_1^1}$ $\to$ $^3\!Q^3_{1/2}$ by addition of one electron 
of spin  $\downarrow$ has the weight  $|Z_{22\downarrow}|^2 \sim 1/3$:
\begin{eqnarray}
\left|
<{^3Q_{1/2}^3}| C^{\dagger}_{22\downarrow} | {^2T_1^1} > 
\right|^2 \!\!\!= 
\left|\langle {^3Q_{1/2}^3}|
C^{\dagger}_{22\downarrow} C^{\dagger}_{11\uparrow} 
C^{\dagger}_{00\uparrow} |0>
\right|^2 \!\!\!= \frac{1}{3}.
\nonumber
\end{eqnarray}
This spectral weight corresponds to the first peak of  fig.~\ref{green3}{\bf c}.
Similarly, the transition  $^2T^1_0$ $\to$  
$^3 Q^3_{1/2}$ has a weight $|Z_{22\uparrow}|^2 \approx$ 
$\left|
<{^3 Q^3_{1/2}}| C^{\dagger}_{22\uparrow} | {^2T^1_0} > 
\right|^2\!\!\!=2/3
$~(fig.~\ref{green3}{\bf d}).
Finally the transition $^2T^1_1$ $\to$ ${^3Q^3_{3/2}}$
has a spectral amplitude $|Z_{22\uparrow}|^2 \approx$
$
\left|
<{^3Q^3_{3/2}}| C^{\dagger}_{22\uparrow} |{^2T^1_1} > 
\right|^2=1
$~(fig.~\ref{green3}{\bf b}).
Comparing these noninteracting estimates  with the exact results,
 the agreement is found to be close.


To study  the linear magnetoconductance we use 
a very general expression for the current derived in the tunneling 
Hamiltonian 
formalism~\cite{meir-jauho}. If  the coupling 
between the left (L)  and right (R) barrier is proportionate, 
the linear conductance can be written 
in terms of the 
transition rates of eq.~\ref{trans}:
\begin{eqnarray}
g_{N,0}(eV_g ) = \left .\frac{dI}{dV} \right |_{N,0} (eV_g )\propto
 \frac{e^2}{\hbar}
 \sum_{\alpha} \int_0^\infty \!\!\!\!\! d\epsilon~
t_{N,0}^{N+1,\alpha} (\epsilon )
 \frac{\partial f}{\partial \epsilon}\:, \nonumber 
\end{eqnarray}
where $f$ is the Fermi function. 
 Just the diagonal terms are  retained in 
eq.~\ref{trans}(what is found to provide  the largest contribution by 
weak coupling, as above mentioned).
 Peaks are present  when the gate voltage with respect to $\mu$ is such 
that $eV_g = \mu ^{N+1,\alpha} _{N,0}$.
 In the absence of crossings,  only the GS  ($\alpha =0 $) 
contributes to the  linear conductance.
Indeed,  other $\alpha$'s   
would correspond to  a nonequilibrium 
condition  for the dot.
 Close to the S-T crossing, the conduction will be the sum of the 
two contributions $g_S$ and $g_T$,
 weighted by the canonical 
equilibrium  probability of the dot being in the singlet 
 or the triplet $^2$GS ($P_S$ and $ P_T^\zeta$ 
respectively, with $\zeta \equiv S_z = \pm 1,0$)~\cite{beenakker}:
\begin{equation}
 \left .\frac{dI}{dV} \right |_{V_{sd}=0}  \!\!\!\!\! (eV_g )=
P_{S} \cdot {g_S} + \sum _{\zeta =\pm 1,0 } 
P_T^\zeta  \cdot{g_T^\zeta} 
\label{fin} 
\end{equation} 
\begin{figure}
\centering
\includegraphics*[width=\columnwidth]{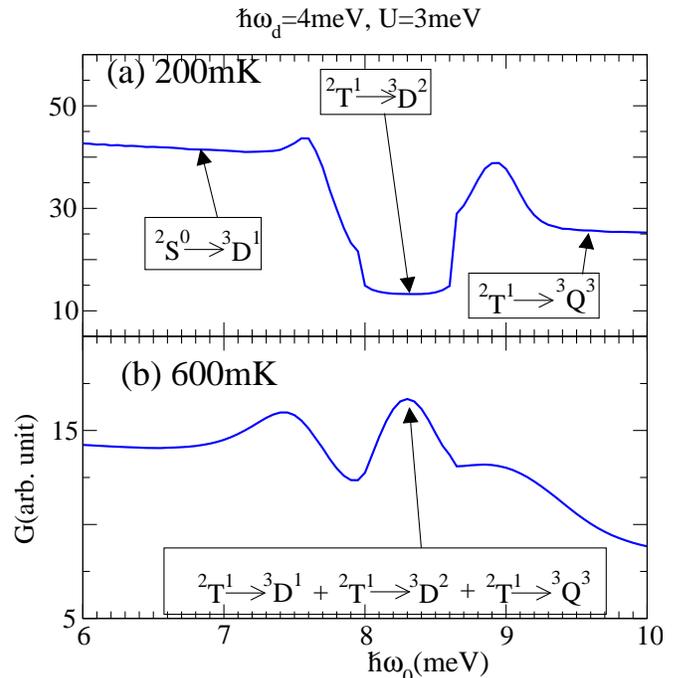}
\caption{Evolution of the height of the conductance peak as a function of
$\hbar \omega_c$.({\bf a}) At a temperature of 200~mK, a dip is clearly visible
in the range $\hbar\omega_c= $ 8-8.8~meV.
({\bf b}) When the temperature increases, 
the dip leaves place to a peak which is due to the contributions 
of the $^2T^1 \rightarrow {^3D^1}$ and $^2T^1 \rightarrow {^3Q^3}$ transitions.} 
\label{kelv3}
\end{figure} 
In fig.~\ref{kelv3}, 
the  maximum of the linear magnetoconductance calculated using 
eq.~\ref{fin} is shown 
at increasing magnetic fields for two  values of the temperature 
(fig.~3{\bf a}: $T=200\mK$, fig.~3{\bf b}: $T=600\mK$). At the lowest 
temperature
it displays sharp features 
close to $B_{ST}$ and a marked dip between $B_{ST}$ and $B_{DQ}$.
Conductance is at a 
maximum when the $^2$GS is a singlet while the $^3$GS is a doublet. 
The same happens at $B> B_{DQ}$  when the $^2$GS is a triplet  while  
the $^3$GS is a quadruplet.  In the latter case, there are 
six tunneling channels
corresponding to  $^2$GS having $S_z = \pm 1,0$ and addition of one extra
electron with up or down spin.  Each of them contributes $ 1/6$  to the total
conductance.  Between $B_{ST}$ and $B_{DQ}$, however, the $^3$GS
  is still a doublet, what inhibits two of the possible tunneling 
channels. 

Increasing temperature (fig.~\ref{kelv3}{\bf b}), 
there is some overlap between the 
peaks due to 
the two doublet states $^3D^1$ and $^3D^2$ and the state $^3Q^3$ as well, 
 which produces  
larger conduction in the intermediate region between $B_{ST}$ and 
$B_{DQ}$ and a new  peak appears.
This eventually washes out  the dip.  

\section{Closing remarks}
 
The  peak height in  Coulomb blockade oscillations 
versus gate voltage $V_g$   is governed by the 
 $^NGS \to {^{N+1}GS}$ transition rate. Using the S-T transition for 
the  $N=2$ particle dot as an example, we have shown that, in 
the neighborhood of $B_{ST}$,
crossing of levels are to be expected for the $N+1$ particle dot as well.
This feature seems to 
be general, as seen from the dot spectrum with $N=4$ and $N=5$.
Increasing the ratio of the Coulomb energy $U$ to the level 
spacing $\hbar \omega_d$ due to the confinement potential, 
moves these crossings to 
lower magnetic fields.

We have calculated the transition rates for addition 
of one electron to the dot with  $N=2$ 
in an orbital with quantum numbers $nm\sigma $.
 They are non-zero  at the discrete energies
  $\epsilon = \mu _{N,0}^{N+1,\alpha}$.
  While the single particle  overlap 
$\Gamma (\epsilon) $ plays a minor role, transition rates 
are  determined in weak coupling   by  the spectral amplitude for the isolated
dot, which 
we obtain by exact diagonalization of a 2D system  with parabolic 
confinement and  Coulomb interaction between the electrons.

Previous numerical work on the spectral weight of a few electron dot was 
mostly concerned with the total quantity $ w _{\alpha} 
= \sum _{nm\sigma} |Z_{nm\sigma}
^{N+1,\alpha} |^2 $(see \cite{pfann} for $B=0$ and \cite{palacios} for 
$\alpha =0$ at various $B$), in which the spectral weights are summed 
over all possible 
orbitals for the added electron.
  They  showed that at definite energies its value could be strongly 
reduced with respect to the noninteracting case, due to correlations.
However, in a vertical configuration for tunneling, with
 $B$ in the direction of the current, 
transition rates for the dot from the $^N$GS 
to the $^{N+1}$GS, or closely lying excited states at $N+1$, 
require
 conservation of the total angular momentum $M$ in the direction of the 
field, as well as  of the total spin component
 $S_z$. 

The quantity $w_\alpha$ is then of little use, if conservation of the
quantum numbers  mentioned  before has to be enforced. 
We have shown that each of the contribution to $w_\alpha$, i.e. each of
 the spectral amplitudes, is strongly dependent on the quantum numbers
 which label the dot states and on the single particle orbital of the
 electron tunneling onto the dot.

The spin selection rule  we have been discussing in Section II 
arises from 
conservation of $S_z$ in a vertical geometry. 
The absence of the quasiparticle (QP) peak for addition of an electron 
of a given spin shows that there are some 
tunneling processes that cannot contribute to the conduction.
This produces  a  dip in the linear 
magnetoconductance, in the range of magnetic field values for which the 
$^2GS \to {^3GS}$ transition is between a triplet $(S=1)$ and a doublet state
$(S=1/2)$.   

As long as the Zeeman spin splitting is ignored, this mechanism is 
not spin sensitive, because $T_{\zeta}$ with $\zeta = 0,\pm 1 $ are 
degenerate in energy. However, if Zeeman spin splitting 
becomes  relevant, 
 the $^3Q^3_{3/2}$ state could become  lower in energy than the 
$^2D^2_{1/2}$ and 
the range of magnetic field values between $B_{ST}$ and $B_{DQ}$,
in which the dip in the conductance occurs,
could disappear. 

Increased correlation, by distributing weight out of the QP peak to 
nearby energies, could reduce the effect.
This is not the case,however,  because  we find that 
much larger $U$ values do not 
change fig.~\ref{green3} 
qualitatively. Instead, they shift the magnetic field at which the crossings 
take place, and, of course, increase the energy differences between the levels.
We do not find a sizeable spreading of the spectral weight to higher 
energies for $U$ values as large as $U=$ 15~meV. This could be partly due to 
the limited number of orbitals used in the single particle basis, however. 
 
We acknowledge useful discussions with A.~Angelucci, M.Di~Stasio, G.~Faini,
R.~Haug, J.~Weis. One of us (A.T.) is grateful to G.~Sch\"on for the 
hospitality at 
the Institut f\"ur Festk\"orperphysik, Universit\"at Karlsruhe,
 where this work 
was written. Work partially supported by INFM (Pra97-QTMD) and 
by EEC with TMR project, contract FMRX-CT98-0180.


\begin{table}[!b]
\centering
\begin{tabular}{ll} 
 ~~~~~level & notation \\ \hline
N=2,~~S=0, M=0~~~~~(singlet) & $^2S^0$ \\
N=2,~~S=1, $S_z$, M=1 (triplet) & $^2T^1_{S_z}$ \\
N=3,~~S=1/2, $S_z$, M (doublet) &    $^3D^M_{S_z}$ \\
N=3,~~S=3/2, $S_z$, M=3~~(quadruplet) & $^3Q^3_{S_z}$  \\ 
\end{tabular}
\caption{Notations for the discrete levels of the QD used in the text.}
\label{notation}
\end{table}


\begin{thebibliography}{99}
\bibitem{nota}
The contribution of the orbital with $n=m$ is the most important one.
Indeed, the spectral weight 
corresponding to $n>m$ is usually very dispersed among  many energies, 
except when the quasiparticle (QP) peak survives intact\cite{pfann},
 but the excitation of the center 
of mass variable takes simultaneously place. This implies that it is 
located at higher energies anyhow.
\end{thebibliography}
\end{document}